\documentclass{aa}
\usepackage{graphicx}
\def\oxy{{\rm O}}
\def\iras{{\it IRAS}}
\def\HI{H {\sc i}}
\def\gmI{\gamma_{\rm Ia}}
\def\gmII{\gamma_{\rm II}}
\begin{document}
\title{Dust-to-gas ratio and star formation history of blue
compact dwarf galaxies}
\author{Hiroyuki Hirashita \inst{1}
\fnmsep\thanks{Research Fellow of the Japan Society for the Promotion of
Science.},
Yuka Y. Tajiri \inst{2},
         \and
Hideyuki Kamaya \inst{2}
\fnmsep\thanks{Visiting Academics at Department of Physics, University
of Oxford, Keble Road, Oxford OX1 3RH, UK (1st March -- 31st December
2001)}
}
\offprints{H. Hirashita}
\institute{Osservatorio Astrofisico di Arcetri, Largo E. Fermi, 5,
50125 Firenze, Italy\\
\email{irasita@arcetri.astro.it}
        \and
Department of Astronomy, Faculty of Science, Kyoto University,
Sakyo-ku, Kyoto 606-8502, Japan\\
\email{tajiri, kamaya@kusastro.kyoto-u.ac.jp}
}
\date{Received 25 July 2001 / accepted 29 March 2002}
\abstract{
This paper investigates the origin of the observed large variety in
dust-to-gas ratio, $\cal{D}$, among blue compact dwarf galaxies
(BCDs). By applying our chemical evolution model, we find that
the dust destruction can largely suppress the dust-to-gas ratio
when the metallicity
of a BCD reaches $12+\log{\rm (O/H)}\sim 8$, i.e., a typical
metallicity level of BCDs. We also show that dust-to-gas ratio is
largely varied owing to the change of dust destruction efficiency
that has two effects: (i) a
significant contribution of Type Ia supernovae to total supernova
rate; (ii) variation of gas mass contained in a star-forming region.
While mass loss from BCDs was previously thought to be the major cause
for the variance of $\cal{D}$, we suggest that the other two effects are
also important.
We finally discuss the intermittent star formation history, which
naturally explains the large dispersion of dust-to-gas ratio among
BCDs.
 }
\titlerunning{Dust-to-gas ratio of BCD}
\authorrunning{H.Hirashita, Y.Y.Tajiri, and H.Kamaya}
\maketitle
%


\section{Introduction}

Dust grains absorb stellar ultraviolet--optical light and emit
far-infrared (FIR) light, thereby affecting the spectral
energy distributions of galaxies (e.g., Takagi et al.\
\cite{takagi99}). Since the spectral energy distribution is frequently
analysed to infer the star formation history (SFH) that provides us
with a key to understand the evolutionary history of galaxies, the
research on the origin of the interstellar dust is important to
this issue. A key quantity concerning grains is the dust-to-gas
mass ratio, $\cal{D}$. Oort \& van
de Hulst (\cite{oort46}) have observationally shown a strong
correlation between the densities of gas and dust. This indicates that
dust traces dense environments, which should be rich in heavy elements. 
Thus, it is worth examining the evolution of ${\cal D}$ in the context
of the chemical evolution of galaxies.

The condensation of heavy elements is an important process for the
formation of grains. One of the environments where condensation
takes place is the atmosphere of a cool giant star (Hoyle \&
Wickramasinghe \cite{hoyle63}). High-dispersion spectroscopic
observation of C$_2$ in post AGB stars may indicate the condensation
process on C$_2$ as a dust kernel (Crawford \& Barlow 2000; Kameswara
Rao \& Lambert 2000).  Another environment for condensation is a
supernova (SN).  Dwek \& Scalo (\cite{dwek80}) have shown that SNe can
be the dominant source of dust grains.  This is partly because of a
rich metal content in SNe. Indeed, a significant amount of dust is
observed within hot SN remnants (Dwek et al.\ \cite{dwek83};
Moseley et al.\ 1989; Kozasa et al.\ 1989) and theoretical work by
Todini \& Ferrara (\cite{todini01}) has explained some principal
features of dust formation in SN 1987A.
Hirashita (\cite{hirashita99a}) has shown that observational dust
amounts suggest that $\sim 10\%$ (the fraction in mass) of the heavy
elements ejected from stars condense into dust grains. Since his model
prediction reproduces the observed trend between dust-to-gas
ratio and metallicity of dwarf galaxies, we adopt this fraction for
dust condensation.

The processes of dust destruction should also be considered. As shown
in Dwek \& Scalo (\cite{dwek80}), dust grains are not only made from
heavy elements but also destroyed in SN shocks (see also McKee
\cite{mckee89}; Jones et al.\ 1996) in a cycle of the birth and death
of stars.  In short, ${\cal D}$ of a galaxy reflects its SFH via the
regulation of dust formation and destruction. To investigate what
determines the value of $\cal{D}$, investigation of a star-forming
galaxy is the most
interesting since SNe are expected to affect most largely their dust
amount.

One of the observational features which give us a key to understand
the regulation of $\cal{D}$ is the variance of $\cal{D}$ itself (e.g.,
Lisenfeld \& Ferrara \cite{lisenfeld98}, hereafter LF98).  In this
paper we examine the variance of $\cal{D}$ among galaxies.
According to LF98, there is a large variance in ${\cal D}$ for a
sample of blue compact dwarf galaxies (BCDs). By definition, galaxies
categorised as BCD more or less show active star-forming activity.
Therefore, dust formation and destruction are expected to occur in
BCDs. Since grains are composed of heavy elements, application of the
theory of galactic chemical evolution is useful and interesting.
Then, we particularly consider the variance of $\cal{D}$ in BCDs by
using the chemical evolution model developed by LF98 and Hirashita
(\cite{hirashita99b}, hereafter H99).

LF98 applied a chemical evolution model to explain $\cal{D}$ of dwarf
galaxies including BCDs. They have suggested that if the dust-to-gas
ratio in outflow (galactic wind) is different from that in the
interstellar medium (ISM), the large dispersion of $\cal{D}$ of dwarf
galaxies can be explained. In this scenario, a significant outflow of
gas is indispensable to explain the observed variance of dust-to-gas
ratio of BCDs. However, is the variance of $\cal{D}$ determined simply
by mass outflow? In this paper, after further investigation in the
framework of LF98, we answer this question and point out that other
factors are also important to explain the value and variance of
$\cal{D}$.

Indeed, a recent analysis by Tajiri \& Kamaya (\cite{tajiri02}) has
suggested that outflow is not so efficient for BCDs. They estimated
the current
momentum supply from SNe by using H$\alpha$ luminosity, and concluded
that the supplied momentum is not sufficient to blow away the
\HI\ envelopes surrounding star-forming regions and that BCDs do not
currently suffer significant mass loss.  Legrand et al. (2001) also
suggested that low density halos around BCDs can be an obstacle for
the ISM to escape from the galaxies themselves.  Thus, it
is worth examining mechanisms other than mass outflow.  Since Tajiri
\& Kamaya (\cite{tajiri02}) adopted a sample in Sage et al.\
(\cite{sage92}), we also use the sample.  Moreover, the sample in LF98
is also included because this paper is an extended study of LF98.

This paper is organised as follows. First, in Sect.\ \ref{sec:model}
we explain the model that describes the evolution of dust content in
a galaxy. In Sect.\ \ref{sec:detail_beta}, we consider dust
destruction, which is the most important process in this paper. Then,
in Sect.\ \ref{sec:result}, model predictions are presented in
comparison with observations. In Sect.\ \ref{sec:discussion}, we
discuss the results and propose a physical mechanism that can explain
the observations. Finally, we summarise the contents of this paper.

\section{Model description}\label{sec:model}
 
In order to consider the dust formation and destruction, we analyse
the dust-to-gas ratio along with the chemical evolution model
by H99. The model is based on Eales \& Edmunds
(\cite{eales96}), LF98 and Dwek (\cite{dwek98}). In our
model, we do not need to model any SFH. This has an advantage in
considering the dust-to-gas ratio of BCDs since the SFH of a dwarf
galaxy is generally complex (e.g., Grebel \cite{grebel01}) and is
difficult to model.  We focus especially on dust destruction,
because LF98 have not fully considered it. As we see later, our model
is a powerful tool to know the metallicity level where dust
destruction becomes effective enough to suppress the dust-to-gas
ratio.

\subsection{Brief review of the model}

In order to investigate dust content in a galaxy, H99 has
established a set of model equations describing dust formation and
destruction processes. In H99, a galaxy is treated as one zone to
focus on the quantities averaged over the whole galaxy. The galaxy
is assumed to be a closed system; that is, mass inflow and outflow
are not considered. If the metallicity of the infalling material is
zero or much lower than that of the ISM in the
galaxy, the relation between dust-to-gas ratio and metallicity, with
which we will be concerned in this paper, is not altered by infall
(Edmunds \cite{edmunds01}; Hirashita \cite{hirashita01}). This is
because the infall dilutes
both metallicity and dust-to-gas ratio at almost the same rate. Our
model does not include the effect of outflow, and this is different
to LF98, in which outflow is essential to explain the observed
variance of the dust-to-gas ratio in BCDs. Since Tajiri \& Kamaya
(\cite{tajiri02}) and Legrand et al. (2001) have suggested that
outflow is not efficient for BCDs, it is worth examining a case of no
outflow.  Indeed, we present another clear possibility to explain the
large scatter of $\cal{D}$ among BCDs later. 

The model equations in H99 (see the paper for details; see also
LF98) describe the evolution of total gas mass ($M_{\rm g}$), the
total mass of metals (both in gas and dust phases) labeled as $i$
($M_{i}$; $i=\oxy$, C, Fe, etc.), and the mass of metal $i$ in a dust
phase ($M_{{\rm d,}\, i}$). We neglect dust growth in clouds, since
Hirashita (\cite{hirashita99a}) has shown that this process in
low-metallicity systems such as dwarf galaxies is much less efficient
than the formation of dust around stars. Then, we adopt an
instantaneous recycling approximation as in LF98 and H99 according to
the formalism in Tinsley (\cite{tinsley80}): Stars less massive than
$m_t$ (present turn-off mass set to be 1 $M_\odot$) live forever and
the others die instantaneously.

\subsection{Solution of the model}

Dust-to-gas ratio and metallicity of galaxies are observationally
known to correlate with each other (e.g., Issa et al.\
\cite{issa90}). This relation has recently been used as a test for
chemical evolution models including dust formation and destruction
(LF98; H99; Hirashita \cite{hirashita99a}; Edmunds \cite{edmunds01}).
The model by H99 reduces the following differential equation:
\begin{eqnarray}
{\cal Y}_i\frac{d{\cal D}_i}{dX_i}= f_{{\rm in},i}
({\cal R}X_i+{\cal Y}_i)-({\cal R}+\beta_{\rm SN}){\cal D}_i\, ,
\label{eq:difeq1}
\end{eqnarray}
where ${\cal D}_i$ is the mass fraction of an element $i$ locked up
in dust (i.e., ${\cal D}_i\equiv M_{{\rm d},\, i}/M_{\rm g}$); $X_i$
is the mass fraction of an element $i$
(i.e., $X_i\equiv M_i/M_{\rm g}$);
$f_{{\rm in},\, i}$ quantifies what fraction of an element $i$
ejected from stars condenses into dust grains;
$\beta_{\rm SN}$ is ``dust destruction efficiency,'' which we explain
later; ${\cal R}$ is the
fraction of stellar mass subsequently returned to the
interstellar space, and ${\cal Y}_i$ is the mass fraction of an
element $i$ newly produced and ejected by stars.\footnote{${\cal R}=R$
and ${\cal Y}_i=y(1-R)$ for the notation in LF98. These two parameters
are calculated in the formalism of the instantaneous recycling
approximation.} The definition of $\beta_{\rm SN}$ is as follows:
\begin{eqnarray}
\beta_{\rm SN}=\frac{M_{\rm g}}{\tau_{\rm SN}\psi}\, ,
\label{eq:beta}
\end{eqnarray}
where $\tau_{\rm SN}$ is the timescale of dust destruction by SN
shocks (Eq.\ \ref{eq:sweeping_time}), and $\psi$ is the star
formation rate (SFR). Hereafter, $\beta_{\rm SN}$ is called
``dust destruction
efficiency'', because it is inversely proportional to the timescale
of dust destruction ($\tau_{\rm SN}$). This is the most important
parameter in this paper.

When all the quantities except $X_i$ and ${\cal D}_i$ are constant
in time, the
analytical solution obtained by LF98 is applicable. With our
notations, it is rewritten as
\begin{eqnarray}
{\cal D}_i(X_i)=\frac{b}{a}X_i+(1-e^{-aX_i})\left(\frac{c}{a}-
\frac{b}{a^2}\right)\, ,\label{eq:solution}
\end{eqnarray}
where
$a \equiv  ({\cal R}+\beta_{\rm SN})/ {{\cal Y}_i} $, 
$b \equiv  {f_{{\rm in},\, i}{\cal R}} / {{\cal Y}_i} $, and 
$c \equiv  f_{{\rm in},\, i}$.

Here, we select oxygen as a traced element (i.e., $i=\oxy$) according
to LF98, because
(i) most of the oxygen is produced by massive stars (Type II SNe and
their progenitors), (ii) oxygen is one of the main constituents of
dust grains, and (iii) the common tracer for the metal abundance in
BCDs is an oxygen emission line. The first item (i) means that an
instantaneous recycling approximation may be reasonable for the
investigation of oxygen abundances, since the generation of oxygen is
a massive-star-weighted phenomenon. In other words, results are
insensitive to the value of $m_t$. Following H99, we adopt
$({\cal R},\,{\cal Y}_\oxy )=(0.32,\, 7.2\times 10^{-3})$, which
are consistent with the relation between dust-to-gas ratio and
metallicity of nearby galaxies including our BCD sample.

\section{Destruction efficiency of dust}\label{sec:detail_beta}

In this paper, we discuss the variance of dust-to-gas ratio of a BCD
sample in terms of the variation of $\beta_{\rm SN}$ defined in
Eq.\ (\ref{eq:beta}). We estimate $\beta_{\rm SN}$ based on McKee
(\cite{mckee89}) and LF98, while we also address their differences.

We assume that gas is divided into two components: gas in the
star-forming region and that in the \HI\ envelope. Such an envelope
is generally observed around a star-forming region of a BCD (e.g.,
van Zee et al.\ \cite{vanzee98}). We denote the gas mass fraction in
the star-forming region as $X_{\rm SF}$ and that in the \HI\ envelope
as $X_{\rm HI}$ (i.e., $X_{\rm SF}+X_{\rm HI}=1$). We distinguish the
two regions for the comparison with the \iras\ sample
(Sect.\ \ref{subsec:obs}). \iras\ FIR bands are sensitive to dust
hotter than about 25 K. Such ``warm'' dust exists in star-forming
regions, not in
\HI\ envelopes.  Calzetti et al.\ (\cite{calzetti95}) have also shown
by using the \iras\ sample of actively star-forming galaxies that 70\%
of the FIR flux comes from such a warm component of dust.  Thus, dust
mass derived from the \iras\ observation of a BCD is considered to
trace the dust in the star-forming region.  This suggests that it is
useful for us to consider dust contained in star-forming regions as
long as we are interested in the comparison of our result with the
\iras\ observations.

Thus, we estimate $\tau_{\rm SN}$ and $\beta_{\rm SN}$ in star-forming
regions.  Gas mass accelerated to a velocity of $v_{\rm s}$ by a SN,
$M_{\rm s}(v_{\rm s})$, is estimated as
\begin{eqnarray}
M_{\rm s}(v_{\rm s})=6800\frac{E_{51}}{v_{\rm s7}^2}~M_\odot\, ,
\end{eqnarray}
where $E_{51}$ is energy released by a SN in units of $10^{51}$ erg
and $v_{\rm s7}$ is $v_{\rm s}$ in units of $10^7$ cm s$^{-1}$ (McKee
1989).  In swept ISM, dust is not fully destroyed. Thus, the fraction
of destroyed dust, $\epsilon$ ($\sim 0.1$; McKee \cite{mckee89}),
should be multiplied. Since we are interested in mass swept by
multiple SNe in a star-forming region, SN rate must be considered,
which is denoted as $\gamma$. Then, $\tau_{\rm SN}$ is expressed as
\begin{eqnarray}
\tau_{\rm SN} =
\frac{M_{\rm g}X_{\rm SF}}
{\epsilon \gamma M_{\rm s}(100~{\rm km~s}^{-1})}\, .
\label{eq:sweeping_time}
\end{eqnarray}
Here, we substitute $v_{\rm s}$ with the threshold velocity for dust
destruction (100 km s$^{-1}$; McKee 1989).  Comparing equations
(\ref{eq:beta}) and (\ref{eq:sweeping_time}), we obtain
\begin{eqnarray}
\beta_{\rm SN}=\epsilon M_{\rm s}(100~{\rm km~s}^{-1})\,
\frac{\gamma}{\psi}\,\frac{1}{X_{\rm SF}}\, .\label{eq:beta_SN}
\end{eqnarray}

We should consider both Type Ia and II SNe in estimating $\gamma$.  Type
II SNe occur ``soon'' after a star formation, because the progenitors of
Type II SNe have lifetimes much shorter than the age of the universe. On
the other hand, Type Ia SNe, whose progenitors are low-mass stars, occur
about 1 Gyr after a star formation.  Thus, $\gamma$ and $\beta_{\rm SN}$
should be sensitive to the SFH. We express $\gamma$ as
\begin{eqnarray}
\gamma =\gmI +\gmII\, ,
\end{eqnarray}
where $\gmI$ and $\gmII$ are the rate of Type Ia and Type II SNe,
respectively. We note that LF98 estimated $\beta_{\rm SN}$ to be 5.
But, for example, the fraction between $\gmI$ and $\gmII$ changes
according to SFH of BCDs.
In the next section, we examine the relation between dust-to-gas
ratio and metallicity for various $\beta_{\rm SN}$, whose probable
range is constrained.

\section{Sample and result}\label{sec:result}

\subsection{Observational samples}\label{subsec:obs}

We compare the model calculation in Eq.\ (\ref{eq:solution}) with the
observed values of dust-to-gas ratio and metallicity of BCDs.  We
adopt the data in Table 2 of LF98 and Table 1 of Sage et al.\
(\cite{sage92}).  The latter sample has been adopted by Tajiri \&
Kamaya (\cite{tajiri02}).  We select BCDs whose 21-cm \HI\ emission
and FIR dust emission are both detected. The data are summarised in
Table \ref{tab:sample}.

  \begin{table*}
     \caption[]{Data of Blue Compact Dwarf Sample}
        \label{tab:sample}
    $$ 
        \begin{array}{p{0.3\linewidth}ccccccc}
           \hline
           \noalign{\smallskip}
           Object & D & S_{60} & S_{100} & \log M_{\rm HI} &
           \log M_{\rm d}^{IRAS} & 12+(\mathrm{O/H}) &\mathrm{ref.^a} \\
                  & \mathrm{(Mpc)} & \mathrm{(Jy)} & \mathrm{(Jy)}
                  & [M_\odot] & [M_\odot] & & \\
           \noalign{\smallskip}
           \hline
           \noalign{\smallskip}
           II Zw 40 & 9.1 & 6.6  & 5.8  & 8.30 & 4.55 & 8.15 & \mathrm{S92} \\
           Haro 2 &  20.3 & 4.8  & 5.5  & 8.68 & 5.39 & 8.4  & \mathrm{S92} \\
           Haro 3 &  13.7 & 5.2  & 6.7  & 8.76 & 5.21 & 8.3  & \mathrm{S92} \\
           UM 439 &  12.6 & 0.39 & 1.2  & 8.23 & 4.96 & 7.98 & \mathrm{S92} \\
           UM 462 &  11.9 & 0.99 & 1.1  & 8.15 & 4.21 & 7.89 & \mathrm{S92} \\
           UM 465 &  13.2 & 0.99 & 1.3  & 7.71 & 4.48 & 8.9  & \mathrm{S92} \\
           UM 533 &  10.4 & 0.51 & 0.54 & 7.76 & 3.75 & 8.10 & \mathrm{S92} \\
           UM 448 &  6.00 & 4.14 & 4.32 & 7.54 & 4.16 & 8.08 & \mathrm{LF98} \\
           IC 3258 & 21.2 & 0.490 & 0.970 & 8.55 & 5.03 & 8.44 &\mathrm{LF98}\\
           Mrk 7 & 42.3 & 0.480 & 0.970 & 9.56 & 5.64 & 8.54 & \mathrm{LF98} \\
           Mrk 33 & 21.6  & 4.68 & 5.30 & 8.77 & 5.41 & 8.40 & \mathrm{LF98} \\
           Mrk 35 & 14.5  & 4.95 & 6.74 & 8.73 & 5.29 & 8.30 & \mathrm{LF98} \\
           Mrk 450 & 12.1 & 0.480 & 0.820 & 7.77 & 4.37 & 8.21&\mathrm{LF98} \\
           NGC 4670 & 12.1 & 2.63 & 4.47 & 8.52 & 5.10 & 8.30 &\mathrm{LF98} \\
           NGC 4861 & 12.9 & 1.97 & 2.26 & 9.13 & 4.60 & 8.08 &\mathrm{LF98} \\
           II Zw 70 & 17.6 & 0.710 & 1.24 & 8.56 & 4.89 & 8.11&\mathrm{LF98} \\
           \noalign{\smallskip}
           \hline
        \end{array}
    $$ 
\begin{list}{}{}
\item[$^{\mathrm{a}}$] S92 --- Sage et al.\ (1992); LF98 ---
  Lisenfeld \& Ferrara (1998)
\end{list}
  \end{table*}

The observational dust-to-gas ratio in a star-forming region should
be
\begin{eqnarray}
{\cal D}^{\rm obs}\equiv M_{\rm d}^{IRAS}/(M_{\rm HI}X_{\rm SF}) \, ,
\end{eqnarray}
where $M_{\rm d}^{IRAS}$ and $M_{\rm HI}$ are dust mass determined
from the \iras\ observation and the total \HI\ gas determined from the
21-cm observations.  $M_{\rm HI}$ traces \HI\ gas in both the
star-forming region and the envelope of a galaxy.  On the other hand,
$M_{\rm d}^{IRAS}$ traces the amount of dust in the star-forming
region, because the \iras\ is sensitive to high-temperature ($\ga 25$
K) dust.  In the following two paragraphs, we describe the two
quantities further.

LF98 divided the observed \HI\ mass by a factor of 2 to obtain the
gas mass in only the star-forming region. In other words, LF98
assumed that $X_{\rm SF}$ is 0.5.  However,
$X_{\rm SF}$ is hardly constrained
observationally. In this paper we assume $X_{\rm SF}=1$
for the observational sample to
obtain the first result. As we discuss in
Sect.\ \ref{sec:discussion}, the variation of $X_{\rm SF}$
also contributes to the variation of $\beta_{\rm SN}$. Therefore,
$X_{\rm SF}$ affects our analysis in the following two ways:
\begin{enumerate}
\item Observationally, the estimate of dust-to-gas ratio in a
star-forming region is affected by the value of $X_{\rm SF}$.
\item Theoretically, the value of $X_{\rm SF}$ affects the value
of $\beta_{\rm SN}$, thus changing the dust-to-gas ratio.
\end{enumerate}
It is difficult to discuss the former quantitatively, because
we need a high-resolution spatial map of \HI\ distribution and
a reasonable observational definition of a star-forming region
on the map. However, we can discuss the latter because
the effect of $X_{\rm SF}$ on $\beta_{\rm SN}$ is well
determined from Eq.\ (\ref{eq:beta_SN}).

The observational dust mass, $M_{\rm d}^{IRAS}$, is derived from the
luminosity densities
at wavelengths of 60 $\mu$m and 100 $\mu$m observed by \iras\ using
Eq.\ (4) of LF98. The \iras\ bands are insensitive to the cold ($\la
20$ K) dust that lies out of star-forming regions. Moreover, since
the dust in a star-forming region suffers destruction by SN
shocks, we should take into account an efficient destruction in the
star-forming region. Therefore, in order to discuss the \iras\ sample
and the selective dust destruction in star-forming regions, we need to
define $\beta_{\rm SN}$ as the {\it dust destruction efficiency}
(Sect.\ \ref{sec:detail_beta}) in star-forming regions as we have done
in Eq.\ (\ref{eq:beta_SN}).

Finally, we hypothesise that the fraction of oxygen contained in
dust grains is
constant for all the sample BCDs. Following H99, we assume the Galactic
composition of the grains:
\begin{eqnarray}
{\cal D}=2.2{\cal D}_\oxy\, .\label{eq:dust_norm}
\end{eqnarray}

\subsection{Conclusion from our model}\label{subsec:result}

\begin{figure}
\includegraphics[width=8cm]{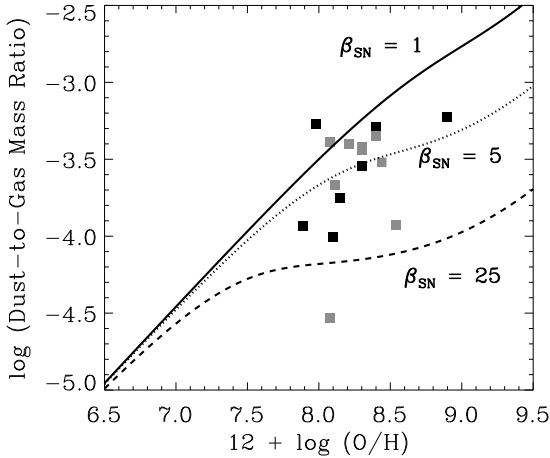}
\caption{Relation between dust-to-gas ratio and
metallicity. The solid, dotted, and dashed lines show the model
results with the dust destruction efficiencies (Eq.\ \ref{eq:beta})
$\beta_{\rm SN}=1$, 5, and 25, respectively. The black and gray
squares show the data for the blue compact dwarf galaxies in
Sage et al.\ (\cite{sage92}) and in
Lisenfeld \& Ferrara (\cite{lisenfeld98}),
respectively.}
\label{fig:result1}
\end{figure}

In Figure \ref{fig:result1}, we show analytical results calculated
according to Eqs.\ (\ref{eq:solution}) and (\ref{eq:dust_norm}) with
$i=\oxy$ for various values of $\beta_{\rm SN}$ as solid line
($\beta_{\rm SN}=1$), dotted line ($\beta_{\rm SN}=5$; the case of
LF98) and dashed line ($\beta_{\rm SN}=25$), respectively.  The black
and gray squares indicate the observational samples in Sage et al.\
(\cite{sage92}) and LF98, respectively. The number ratio of oxygen
atoms to the hydrogen atoms is denoted as (O/H). We convert the mass
fraction of oxygen, $X_\oxy$, to (O/H) for the model prediction,
assuming $\log X_\oxy +10.80=12+\log{\rm (O/H)}$.  In this figure,
$f_{\rm in,\, O}=0.1$ is assumed to concentrate on the variation in
$\beta_{\rm SN}$. LF98 have shown a large variety of $f_{\rm in,\, O}$
to explain the large scatter of the relation between $\cal{D}$ and
$X_\oxy$ in their Figure 7. Since Hirashita
(\cite{hirashita99a}) has shown that $f_{\rm in,\, O}\sim 0.1$
reproduces the observed trend between dust-to-gas ratio and
metallicity, it is worth considering the effect of
$\beta_{\rm SN}$ by setting $f_{\rm in,\, O}=0.1$.

{}From Figure \ref{fig:result1}, we see that the variance in
dust-to-gas ratio is reproduced by an order-of-magnitude variation in
$\beta_{\rm SN}$ (1--25 here). We note that LF98's value ($\beta_{\rm
SN}=5$) is within this range.  Even if outflow does not efficiently
occur in BCDs, we can explain the variance of the dust-to-gas ratio
of BCDs with the various ``destruction efficiency'' of dust,
$\beta_{\rm SN}$, although we can never reject the importance of
outflow in the
framework of this paper. The important point is that we have
demonstrated that dust destruction by SNe can play an important role
in producing the variance of $\cal D$.
The result also indicates that the difference in dust-to-gas ratio
among the three lines becomes clear at an oxygen abundance, $12+\log
({\rm O/H})\sim 8$. Thus, we conclude
\begin{enumerate}
\item that the dust destruction by SNe can vary the dust-to-gas
ratio when the metallicity, $12+\log ({\rm O/H})$, reaches about 8
($\sim 10$\% of the solar metallicity), and 
\item that the large variation of the dust-to-gas ratio among
BCDs is explained by the variation of dust destruction efficiency,
$\beta_{\rm SN}$.
\end{enumerate}

\section{Discussion}\label{sec:discussion}

\subsection{Scenario}

We propose a scenario for large variation of $\beta_{\rm SN}$ for each
BCD along with intermittent SFH.   The
time variability of SFH on a short ($<1$ Gyr) timescale has been
suggested observationally by Searle \& Sargent (\cite{searle72}) and
theoretically by Gerola et al.\ (\cite{gerola80}). Nonlinear
processes in the ISM may also cause an intermittent star formation
(Ikeuchi \cite{ikeuchi88}; Kamaya \& Takeuchi \cite{kamaya97} and
references therein).
Thus, it is pertinent to consider the intermittent SFH of BCDs.

As shown in the following, an intermittent star formation
history leads to the time variation of $\beta_{\rm SN}$, because
$\gamma /\psi$ depends on time. We consider an intermittent SFH: a
starburst whose SFR is $\psi_{\rm burst}$ and an inter-starburst
epoch whose SFR is $\psi_{\rm inter}$. We assume that
$\psi_{\rm burst}=100\psi_{\rm inter}$, for example. Such a
two-orders-of magnitude variation in SFR is proposed theoretically
by Gerola et al.\ (1980) and Kamaya \& Takeuchi (\cite{kamaya97}).
While the starburst is going on, we expect that
almost all the SNe are Type II ($\gamma_{\rm burst}\sim \gmII$,
where $\gamma_{\rm burst}$ is a typical SN rate in the bursting epoch).
However, in the inter-burst epoch, Type Ia SNe can be dominated
($\gamma_{\rm inter}\sim\gmI$, where $\gamma_{\rm inter}$ is a typical
SN rate in the inter-burst). According to
the model by Bradamante et al.\ (\cite{bradamante98}), a given
stellar population releases energy in the form of Type II and Ia
SNe with a ratio of 5:1 (Bradamante et al.\ \cite{bradamante98};
see their Fig.\ 9. This value is essentially determined
by the initial mass function (IMF), and they assumed the Salpeter
IMF with a
stellar mass range from 0.1 to 100 $M_\odot$)\footnote{We assumed
the Salpeter IMF with the range from 0.1--120 $M_\odot$, but
the difference in the higher mass cut has little influence
on the SN rate.}. Since they assumed the
same energy between Type Ia and II SNe, this means that
the number ratio between Type Ia and II SNe is 5:1.
Therefore, we expect that
$\gamma_{\rm inter}\sim\gamma_{\rm burst}/5$. The intermittent
star formation finally predicts that
$\gamma_{\rm inter}/\psi_{\rm inter}\sim 20\gamma_{\rm burst}/
\psi_{\rm burst}$. This means that an intermittent SFH can
cause a 20-times variation in $\beta_{\rm SN}\propto\gamma /\psi$
during a single star formation cycle.

Furthermore, Fig.\ \ref{fig:result1} shows that the value of $\beta_{\rm
SN}$ has little effect on the relation between dust-to-gas ratio and
metallicity for $12+\log{\rm (O/H)} < 8$. Therefore, until the
metallicity level becomes $12+\log{\rm (O/H)}\sim 8$, the relation
between dust-to-gas ratio and metallicity evolves in the same way
whatever the value of $\beta_{\rm SN}$ might be.  On the contrary, the
relation is largely affected by $\beta_{\rm SN}$ if $12+\log{\rm
(O/H)}>8$. Then, we study the response of the relation between
dust-to-gas ratio and metallicity to the
change of $\beta_{\rm SN}$ at $12+\log{\rm (O/H)} \sim 8$, as we are
interested in the intermittent SFH.

First, we shall estimate a typical metallicity increment during a single
star formation epoch of the intermittent SFH. The metallicity
increase during an episode of star formation, $\Delta Z$, can be
estimated by $\Delta Z\sim yM_*/M_{\rm g}$, where $M_*$ is the
mass of stars formed in the episode, and $y$ is a chemical yield.
If the IMF is similar
to that of the Galaxy, $y\sim Z_\odot$ (i.e., $\Delta Z\to Z_\odot$
for $M_*\to M_{\rm g}$). We estimate $M_*$ by multiplying
observed SFR with a duration of an episode of a star formation
activity. Assuming that the SFR is
0.1 $M_\odot~{\rm yr}^{-1}$ and that the duration is $10^7$ yr
(Legrand et al.\ \cite{legrand01}), we obtain
$M_*\sim 10^6~M_\odot$. With typical gas mass
$M_{\rm g}\sim 10^7~M_\odot$, we obtain $\Delta Z\sim 0.1Z_\odot$.
This corresponds, for example, to the metallicity increase
from $12+\log{\rm (O/H)}=8.0$ to 8.2. The model by Bradamante et
al.\ (\cite{bradamante98}) also indicates that one
episode of star formation can result in such a metallicity increment.

\begin{figure}
\includegraphics[width=8cm]{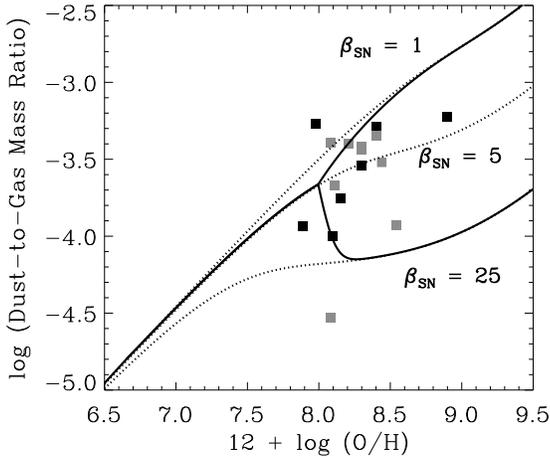}
\caption{Relation between dust-to-gas ratio and
metallicity. The three dotted lines
show the model
results same as Fig.\ \ref{fig:result1}. The squares
are the data points same as Fig.\ \ref{fig:result1}.
The two solid lines represents the result of
the calculations which change $\beta_{\rm SN}$
(from 5 to 1 and from 5 to 25 for upper and lower lines,
respectively).}
\label{fig:change_beta}
\end{figure}

As shown above, the effect of intermittence can be examined by
changing $\beta_{\rm SN}$. In order to examine the effect of time
variation of $\beta_{\rm SN}$,
thus, we calculate the relation between dust-to-gas ratio and
metallicity in the following two cases:
\begin{enumerate}
\item $\beta_{\rm SN}=5$ for $12+\log{\rm (O/H)}\leq 8$ and
$\beta_{\rm SN}=25$ for $12+\log{\rm (O/H)}>8$
\item $\beta_{\rm SN}=5$ for $12+\log{\rm (O/H)}\leq 8$ and
$\beta_{\rm SN}=1$ for $12+\log{\rm (O/H)}>8$
\end{enumerate}
In Figure \ref{fig:change_beta}, we show the result of the two
calculations (two solid lines). The lower and upper branches represent
the cases 1 and 2, respectively. The three dotted lines show
the results as in Fig.\ \ref{fig:result1}.
We see that when the metallicity
$12+\log{\rm (O/H)}$ increases from 8.0 to 8.2, the line rapidly
converges to the two dotted lines, which represent the result for
constant $\beta_{\rm SN}$ (1 and 25, respectively). This rapid
convergence supports the idea that the dust-to-gas ratio varies
widely in response to the time-evolution of $\beta_{\rm SN}$
in an episode of
star formation at the metallicity level of $12+\log{\rm
(O/H)}\sim 8$. Thus, it is possible to determine the variance of
the dust-to-gas ratio among BCDs from the time variation of
$\beta_{\rm SN}$.


\subsection{Age difference}

If the age of BCDs varies, the present turn-off mass of stars is
changed. As a result, the returned fraction of gas (${\cal R}$), the
metal yield (${\cal Y}_i$), and the dust supply from stars
($f_{{\rm in},\,i}$) are effectively different among BCDs. The
effect of varying
${\cal R}$ and ${\cal Y}_i$ on the relation between dust-to-gas ratio
and metallicity has been examined by H99.\footnote{See Fig.\ 3 of the
paper. Although H99 examined ``different IMF'', this difference only
affects the values of ${\cal R}$ and
${\cal Y}_i$. Therefore, his result can be used to see the effect of
varying ${\cal R}$ and ${\cal Y}_i$.} However, the resulting relation
is less sensitive to the two parameters than to $\beta_{\rm SN}$.  The
dependence of $f_{{\rm in},\,i}$ on the turn-off mass can be important
since it largely affects the dust amount in low-metallicity systems
(LF98; Hirashita \cite{hirashita99a}).

Thus, if the BCD sample proves to have a large age variation, we
should reconsider the variation in dust-to-gas ratio with a
time-dependent formulation.  Indeed, we cannot reject the possibility
that some BCDs
are much younger than the cosmic age.  A metal-poor BCD SBS 0335--052
may be younger than $5\times 10^6$ yr (Vanzi et al.\
\cite{vanzi00}). Recently, Hirashita, Hunt, \& Ferrara
(\cite{hirashita02}) have succeeded in explaining the dust amount of SBS
0335--052 with a time-dependent formulation applicable to young
galaxies.

\subsection{Future observations}\label{subsec:future}

The time variation of $X_{\rm SF}$ also leads to the time dependence
of $\beta_{\rm SF}$ (Eq.\ \ref{eq:beta_SN}),
although $X_{\rm SF}$ was assumed as unity (Sect.\ \ref{subsec:obs}).
If the various $X_{\rm SF}$ for the BCD sample is interpreted to
reflect the
time evolution of $X_{\rm SF}$ in each BCD, we can suggest that the
gas mass in a star-forming region should change temporally because of
the mass exchange between the star-forming region and the envelope.
Such a
mass exchange during episodic star formation activity in BCDs is indeed
suggested by e.g., Sait\={o} et al.\ (\cite{saito00}). The
temporal change of $X_{\rm SF}$ is also possible
if the ISM in the star-forming region is consumed for star
formation and locked in stellar remnants like white dwarfs, neutron
stars, and black holes.

In order to constrain $X_{\rm SF}$, we need to observe \HI\ emission
or FIR emission with angular resolution fine enough. The present
typical angular resolution of $1'$ corresponds to 2.9 kpc in physical
size if a galaxy lies at a typical distance of 10 Mpc.
Future large space FIR telescopes such as
{\it Herschel}~\footnote{http://astro.estec.esa.nl/First/} (e.g.,
Pilbratt \cite{pilbratt00}) or {\it
SPICA}~\footnote{http://www.ir.isas.ac.jp/SPICA/index.html} (e.g.,
Nakagawa et al.\ \cite{nakagawa00}) will resolve the star-forming
regions of the BCDs. For example, the Japanese future infrared
satellite {\it SPICA} will have a diameter larger than 3.5 m.  If the
diffraction limit is achieved, the angular resolution becomes $6''$
at the
wavelength of 100 $\mu$m. This corresponds to 290 pc at the distance
of 10 Mpc, and is comparable to or smaller than the half-light
radius of a typical BCD (Marlowe et al.\ \cite{marlowe99}).

\section{Summary}\label{sec:summary}

In this paper, we have analysed the relation between dust-to-gas ratio
and metallicity of BCDs by using our chemical evolution model. We have
focused on the dust destruction process, because this process was not
investigated in LF98. We have shown that the dust destruction
significantly affects the dust-to-gas ratio when the metallicity is
larger than $12+\log{\rm (O/H)} \sim 8$. The intermittent SFH can
explain the large variety of dust-to-gas ratio among BCDs through the
``dust destruction efficiency'' of $\beta_{\rm SN}$.

\begin{acknowledgements}

We thank the anonymous referee for invaluable comments and suggestions
which improved this paper very much.  We thank A. Ferrara and
K. Yoshikawa for useful discussions on dwarf galaxies and excellent
environments for our study. We are also grateful to S. Mineshige and
J. Silk for continuous encouragement. H. H. was supported by the
Research Fellowship of the Japan Society for the Promotion of Science
for Young Scientists.  We fully utilised the NASA's Astrophysics Data
System Abstract Service (ADS).

\end{acknowledgements}




\end{document}